# RELIABILITY HISTORY AND IMPROVEMENTS TO THE ANL 50 MEV H- ACCELERATOR*

L.I. Donley, V.F. Stipp, F. R. Brumwell, G. E. McMichael, ANL, Argonne, IL 60439, USA

*Abstract*

The H- Accelerator consists of a 750 keV Cockcroft Walton preaccelerator and an Alvarez type 50 MeV linac. The accelerator has been in operation since 1961. Since 1981, it has been used as the injector for the Intense Pulsed Neutron Source (IPNS), a national user facility for neutron scattering. The linac delivers about $3.5 \times 10^{12}$ H- ions per pulse, 30 times per second (30 Hz), for multi-turn injection to a 450 MeV Rapid Cycling Synchrotron (RCS). IPNS presently operates about 4,000 hours per year, and operating when scheduled is critical to meeting the needs of the user community. For many years the IPNS injector/RCS has achieved an average reliability of 95%, helped in large part by the preaccelerator/linac which has averaged nearly 99%. To maintain and improve system reliability, records need to show what each subsystem contributes to the total down time. The history of source and linac subsystem reliability, and improvements that have been made to improve reliability, will be described. Plans to maintain or enhance this reliability for at least another ten years of operation, will also be discussed.

## 1 INTRODUCTION

After operating with 200 μs pulses at 0.25 Hz for 15 years as an injector for the Zero Gradient Synchrotron (ZGS), the linac repetition rate was increased to 30 Hz with 70 μs pulses in the mid-1970's when it also became an injector for the Rapid Cycling Synchrotron (RCS). Direct injection to the ZGS ended in 1979 and the RCS, which was originally built to be a booster for the ZGS, became the source of high-energy protons for the Intense Pulsed Neutron Source (IPNS), a national user facility for neutron scattering. Since IPNS began operation in 1981, it has accumulated over 60,000 hours of 30 Hz beam operations. The overall accelerator reliability (beam hours versus scheduled hours) exceeds 93% and for the past 10 years, that number is greater than 95%. Thoughout, the preaccelerator/linac reliability has hovered around 99%, and this has been achieved while operating at roughly 30 times its original duty factor.

The evolution and performance of the preaccelerator and linac was described in a 35[th] anniversary paper presented at the Linac96 Conference [1]. In this present paper, we examine overall system and subsystem reliability and the impact of changes in hardware and software on reliability, and discuss improvements that will be necessary to maintain current high performance for at least another ten years.

The linac rf system was the first linac amplifier built using the 7835 triode. The rf amplifier system was designed and built by Continental Electronics Corporation in the late 1950's and comprises a single 4-stage amplifier system with a rated output of 5 MW and a maximum pulse length of about 200 μs. With the exception of the power supplies, most of which had to be upgraded during the late 1970's when the duty factor was increased, the original configuration and in many cases, original equipment, is still functioning well. In the almost 40 years since startup, seventeen new or rebuilt 7835 tubes have accumulated a total of over 200,000 filament hours in this single station, equivalent to over 23 years of continuous operation.

## 2 PREACCELERATOR

The preaccelerator is a standard 4-stage 750 kV Cockcroft-Walton power supply built by Haefely. There have been no failures of the Haefely high-voltage components in the life of the facility. The original motor generator set supplying 400 Hz ac to the high-voltage rectifiers was replaced by one made in the USA. The original 400 Hz isolated generator used for supplying input power to the ion source enclosure was also replaced with a brushless 60 Hz unit to reduce maintenance. It has been trouble-free for the past 10 years.

The first ion source was a proton source. This was changed to a modified duoplasmatron-type H- source in 1976, and then to the present magnetron-type H- source in 1983. The present source uses cesium to enhance the H- current and produces 45-50 mA, 70 microsecond beam pulses at the required 30 Hz rate. It has been very reliable and typically is dismantled and cleaned every six months, with new electrodes installed about every year. Close control of the cesium flow from the ion source is necessary to keep the spark rate of the 750 kV column low. However, proof of success is the fact that the column has not been dismantled in over 14 years.

Detailed operating logs have been kept since the beginning of operation and for the last five years, the trouble log has been computerized. Accelerator downtime is tracked, rounded to the closest 5 minutes. Over the last five years (20,000 hours of beam operation) there were 143 interruptions of 5 minutes or more charged to the preaccelerator and 120 to the linac. Breakdown by subsystem is given in Tables 1 and 2.

*This work is supported by the US DOE under contract no. W-31-109-ENG-38.

Table 1: Lost beam time in last 20,000 hours of operation due to preaccelerator faults (number of beam interruptions, lost beam time, mean-time-between-failure and mean-time-to-repair).

| Subsystem | # of Faults | Down Time (hours) | MTBF (weeks) | MTTR (hours) |
|---|---|---|---|---|
| Haefely Supply | 27 | 7.2 | 4.4 | 0.3 |
| Chopper | 15 | 4.0 | 7.9 | 0.3 |
| Extractor Supply | 4 | 1.0 | 29.7 | 0.3 |
| Controls | 23 | 11.3 | 5.2 | 0.5 |
| Bouncer Supply | 18 | 10.6 | 6.6 | 0.6 |
| H⁻ Source | 25 | 14.0 | 4.7 | 0.6 |
| Vacuum/Water | 22 | 7.3 | 5.4 | 0.3 |
| Beam Transport | 9 | 7.3 | 13.2 | 0.8 |

# 3 LINAC

The 50 MeV linac is an Alvarez structure, constructed as eleven sections that are bolted together to form a single rf cavity that is 0.94 m in diameter and 33.5 m long. There are dc quadrupole magnets in each of the 124 drift tubes powered by twelve mag-amp type dc supplies. These units are original to the linac and are nearly 40 years old. The change from tube to semiconductor regulators in 1988 cleared up most of the maintenance and failure problems with these systems. The tank, drift tubes and quadrupoles have been very reliable. With the exception of a failed interlock and loss of water flow in the first eight quadrupole magnets in 1988 (in which external soft-solder joints melted and separated but no noticeable internal damage to the magnet coils occurred), these systems have caused no significant downtime. Occasional rf arcing in the linac tank after vacuum work has generally been cured by mild reconditioning, and there is sufficient redundancy of vacuum pumps that failures can be left to the next scheduled maintenance period for repair.

Over the life of the machine, there have been significant changes in the vacuum pumps used on both the preaccelerator and linac. Some of the original 2000 l/s ion pumps are still in use (after many rebuilds to replace the titanium elements and bead-blast internal surfaces). Others have been replaced with cryopumps or turbomolecular pumps to increase pumping speed and allow operation to continue despite small leaks. Vacuum pump upgrades are expected to continue over the life of the facility because improvements in vacuum have positive effects on many of the subsystems.

As is evident from Table 2, the rf transmitter accounts for most of the downtime, and over the past five years almost half the downtime for the preaccelerator/linac combined is caused by two components, the dc blocking capacitors in the triode cavity. These capacitors, which are custom made by Continental Electronics and unique to our transmitter, only became a problem after the change to 30 Hz operation. However, in the past twenty years, the upper capacitor has failed 16 times and the lower 10 times. Although average lifetimes are approximately 25 and 40 operating weeks for the upper and lower capacitors respectively, the range is broad. There was a period from 1987 through 1995 (170 weeks) with no failures, whereas others have failed after only a few weeks. Frequency of failures have been higher over the past few years, possibly due to higher-order modes in the cavity and to difficulties the manufacturer is encountering in maintaining a rebuild capability for such limited-demand items.

Table 2: As Table 1 but for linac faults.

| Subsystem | # of Faults | Down Time (hours) | MTBF (weeks) | MTTR (hours) |
|---|---|---|---|---|
| DT power supplies | 14 | 9.6 | 8.5 | 0.7 |
| RF (capacitors) | 15 | 109.6 | 7.9 | 7.3 |
| RF (triode) | 1 | 4.5 | 118.7 | 4.5 |
| RF (other) | 64 | 34.4 | 1.9 | 0.5 |
| Vacuum/Water | 19 | 11.3 | 6.2 | 0.6 |
| Controls | 7 | 3.1 | 17.0 | 0.4 |

The 7835 triode amplifier tubes used in our linac and at several other accelerator facilities (e.g., LANL, FNAL) were originally developed and manufactured by RCA Corporation. For many years, the only supplier for these tubes has been Burle Industries, a small company set up by some former employees of RCA. Lifetime of the tubes is a significant operational consideration because of the high item cost (of order $150,000 for a new tube and half that if the failed tube can be repaired) and delivery times are 6 to 12 months. Since startup in 1961, we have accumulated over 200,000 filament hours on 17 new or rebuilt tubes. In the early years of the facility, most of the tubes failed during operation because of grid-cathode shorts. Lifetimes varied from 3,000 to 17,000 hours (10,000 hours average) for the eight tubes that developed shorts, and each resulted in downtime to install and condition the new tube. A manufacturing change by RCA about 1970 cured the grid-cathode short problem, and now the tubes last until the cathode degrades. Cathodes degrade gradually, giving several months during which the heater current can be periodically increased to maintain sufficient emission and providing time to schedule the replacement during a maintenance period. Nine tubes have been replaced because of low emission with lifetimes ranging from 9,000 to 20,000 hours (15,000 hours average). The only downtime charged to the tubes in the last five years was a 4.5 hour interruption in 1997, a few days after a blocking-capacitor failure, when the tube developed an external arc (carbon track on the ceramic). The tube was

removed, cleaned and returned to service and proceeded to run trouble-free for another 14,000 hours.

## 4 DISCUSSION

IPNS operation increased about five years ago to its present level of about 25 weeks per year. During operating periods, typically two to three weeks on with one to two weeks off between runs, the facility is expected to run 24 hours a day, seven days a week. External users come for periods of a couple days to a week or more to make measurements on samples requiring individual exposure times that vary from less than an hour to several days. In general, beam interruptions of half an hour or less are relatively insignificant to users unless they are frequent. Even then, the main effect is a lack of throughput proportional to the lost beam (equivalent to operating continuously at reduced proton current on target). Long interruptions, particularly those exceeding eight hours, may mean that the user's samples don't all get exposed or statistics are poor and the user may get ambiguous or uninterpretable data. There have been only five beam interruptions from preaccelerator/linac faults over the last five years that exceeded eight hours and the longest was 20 hours. All were associated with blocking-capacitor failures. The average preaccelerator/linac reliability (percent of scheduled time when protons were delivered to target) was 98.8% and half of the total downtime was associated with blocking-capacitor failures. Downtime of all IPNS accelerator systems over the past 12 years (including the preaccelerator/linac) is shown in Figure 1. Overall, the preaccelerator/linac accounted for about one quarter of the total accelerator downtime.

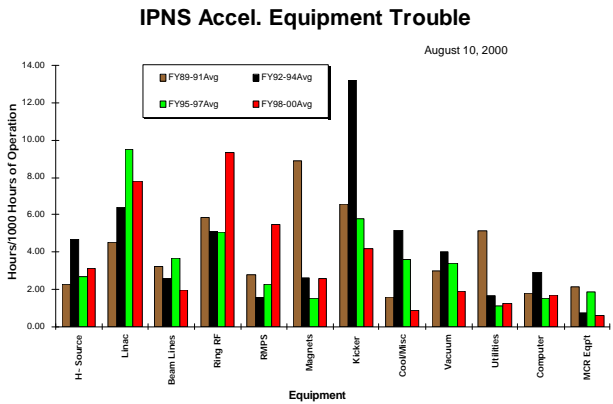

Figure 1. IPNS Accelerator Equipment Trouble Chart

The IPNS accelerators, for over 10 years, have delivered high-energy protons to the spallation target with an average reliability of 95%. Experience has shown that this is a level at which the neutron-scattering users are seldom disappointed. However, on the occasional two-to-three week runs when reliability slips below 90%, the chances of some users not completing their experiments are much increased. Also, the short nature of individual experiments means that one cannot compensate for a lost day of beam by operating at 10% higher current for the next 10 days. Thus for IPNS, mean-time-to-repair is at least as important as total downtime.

The challenge posed five years ago, of increasing operation from the previous level of 15-17 weeks per year to 25 or greater without impacting reliability, has been met. Present plans call for a further increase to about 30 weeks per year. Additional pressures on reliability are that several systems, specifically the high-voltage regulator and pulsed bouncer for the preaccelerator, and the driver stage for the 7835 power triode in the linac rf system, will have to be replaced because the tubes used in these units are no longer available. As it is probable that the new systems will result in an initial increase in downtime, it will be important to aggressively attack the linac rf capacitor problem which now accounts for about half of our total downtime. The pressurized cavities used for the 7835 tubes at FNAL and LANL are a possible solution, but could not be installed in the 8-10 week summer shutdown of our present operating schedule. Therefore we are hoping that our present work with the manufacturer to develop a more robust capacitor, coupled with our work to decrease higher-order modes in the cavity, will provide relief.

## ACKNOWLEDGEMENTS

The authors strongly believe that the reliability of the IPNS accelerator systems is primarily attributable to the designers, subsystem managers, engineers, technicians and operators who have systematically eliminated the weak components and have developed maintenance and operating techniques that provide rapid detection and repair when failures occur. Their response to calls, at any hour of the day or night, is what keeps our mean-time-to-repair manageable.